\documentclass[aps,prl,reprint,groupedaddress]{revtex4-1}

\usepackage{graphicx}
\usepackage{bm}
\usepackage{verbatim}
\usepackage{color}
\usepackage{overpic}
\usepackage{amsmath}
\usepackage[usenames,dvipsnames]{xcolor}

\begin{document}


\title{Focusing and sorting of ellipsoidal magnetic particles in microchannels}


\author{Daiki Matsunaga}
\author{Fanlong Meng}
\author{Andreas Z\"ottl}
\author{Ramin Golestanian}
\author{Julia M. Yeomans}
\email{julia.yeomans@physics.ox.ac.uk}
\affiliation{%
 Rudolf Peierls Centre for Theoretical Physics, University of Oxford\\
 1 Keble Road, Oxford, OX13NP
}%


\date{\today}

\begin{abstract}
We present a simple method to control the position of ellipsoidal magnetic particles in microchannel Poiseuille flow at low Reynolds number using a static uniform magnetic field.
The magnetic field is utilized to pin the particle orientation, and the hydrodynamic interactions between ellipsoids and channel walls allow control of the transverse position of the particles.
We employ a far-field hydrodynamic theory and simulations using the boundary element method and Brownian dynamics to show how magnetic particles can be focussed and segregated by size and shape.
This is of importance for particle manipulation in lab-on-a-chip devices.
\end{abstract}

\pacs{}

\maketitle


Nowadays microscopic lab-on-a-chip devices have become powerful tools to analyse, manipulate and control droplets \cite{Dreyfus2003,Stone2004,Link2004,Beatus2006,Beatus2012}, biological particles \cite{Pamme2007,Bhagat2010,Sajeesh2014} and active colloids \cite{Lindner2014,Das2015,Bechinger2017}.
Different particle types can be separated in microfluidic channels where a steady Poiseuille flow is imposed \cite{Brady1994,Squires2005}.
In particular, positional control along the transverse direction of the channel is desirable in order to transport particles to outlets at different target positions.
Under high Reynolds number flow, inertial forces lead to a migration of particles towards stable positions \cite{Segre1961, DiCarlo2009} which can be manipulated by feedback-control \cite{Prohm2014}.
In contrast, at low Reynolds number, which is the usual regime at the micron scale, spherical and elongated particles cannot achieve net transverse motion in the absence of external forces \cite{Bretherton1962,Goldman1967,Pozrikidis2005a,GuazzelliBook}.
In this work, we focus on this low Reynolds number regime.

In lab-on-a-chip devices, magnetic forces are commonly used to manipulate the position of microscopic particles \cite{Yan2012, Hejazian2015, Huang2017} or artificial microswimmers \cite{Peyer2013, Hamilton2017}.
For example, the segregation of different particle types can be realized by applying an external magnetic field gradient, which essentially acts as a body force \cite{Hejazian2015}.
Although this method can be used to segregate different types of particles, it can not be used to focus particles to a specific transverse target position.
When a uniform magnetic field is applied instead of a gradient field, the particle will only experience a torque but no force, i.e. a uniform field is useful to change the orientation of the particle \cite{Erb2016}, but it is not an intuitive way to achieve translation.
Interestingly, Zhou et al.~\cite{Zhou2017} recently showed that paramagnetic ellipsoidal particles can be focussed to the channel center by applying a static uniform magnetic field perpendicular to the flow.
They managed to achieve net motion away from the wall by breaking the symmetry of cyclic up-down motion \cite{Pozrikidis2005a} of the ellipsoid.

Here we show that the particle position can be controlled not only to the channel center, but to arbitrary target channel positions by using a static uniform magnetic field to pin the orientation of the magnetic particles.
Firstly, we show that the particle will continuously move either towards or away from the wall, purely by hydrodynamic particle-wall interactions.
Secondly, we demonstrate that the ellipsoidal particle can be focused to arbitrary transverse target positions just by a simple manipulation of the magnetic field.

\begin{figure}
    \includegraphics[width=\columnwidth]{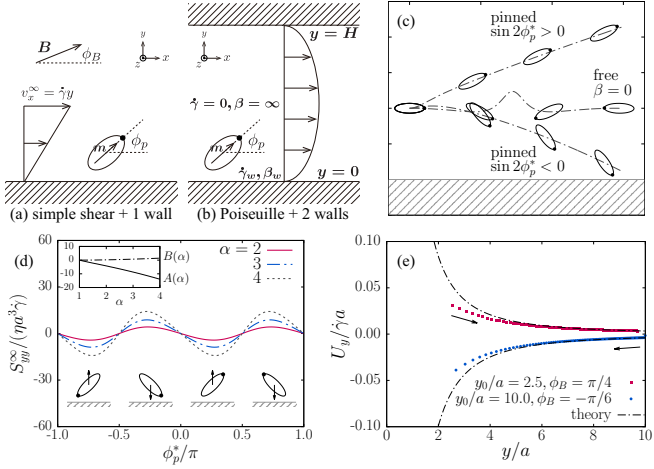}
    \caption{(a)-(b) 2D schematic showing the problem geometry where $\mathbf{m} = (m \cos \phi_p, m \sin \phi_p, 0)$ is the magnetic moment of the ellipsoidal magnet and $\mathbf{B} = (B \cos \phi_B, B \sin \phi_B, 0)$ is the applied uniform magnetic field. Circles on the particle show the direction of the magnetic moment. (c) Schematic of particle movement with different pinned orientations $\phi_p^*$. (d) Stresslet $S_{yy}^\infty$ as a function of pinned orientation $\phi_p^*$. The inset shows the magnitude of the coefficients of Eq.~(\ref{eq:Syy}), and particle schematics show the direction of the transverse movement. (e) Comparison of far-field theory Eq.~(\ref{eq:lift}) and boundary element simulations for the lift velocity $U_y$ of an ellipsoidal particle with $\alpha=3$ and $\beta = 100$. \label{fig:figure1}}
\end{figure}

We consider a permanent magnetic particle with prolate shape of volume $4 \pi a^3/3$, suspended in a Newtonian fluid of viscosity $\eta$ and density $\rho$.
The particle has magnetization $M$, and it is assumed to be neutrally buoyant for simplicity.
It has one semi-axis of length $b_1 = a \alpha^{2/3}$ and two of length $b_2 = a \alpha^{-1/3}$, where $\alpha$ is the particle's aspect ratio $\alpha = b_1/b_2 > 1$.
The particle has a magnetic moment $\mathbf{m} = (m \cos \phi_p, m \sin \phi_p, 0)$ where $m = 4 \pi a^3 M/3$ is the magnetic moment parallel to the particle's major axis and $\phi_p$ is the particle orientation angle [Fig.~\ref{fig:figure1}(a)-(b)].
The particle is initially placed a distance $y_0$ away from a infinite plane wall located at $y=0$, and it experiences a magnetic torque $\bm{T}_m = \bm{m} \times \bm{B}$ due to a uniform external field $\bm{B}$ applied to the whole domain.
We assume that $\bm{B}$ is oriented in ($x, y$)-plane, $\bm{B} = (B \cos \phi_B, B \sin \phi_B, 0)$, where $B$ is the strength and $\phi_B$ the orientation of the field which are both kept constant [Fig.~\ref{fig:figure1}(a)].
Note that we only consider in-plane motion of the particle in this paper, because a strong magnetic field will orient the major axis of the particle in-plane \cite{Almog1995}.
We introduce a non-dimensional parameter $\beta$ that describes the strength of the magnetic torque compared to the hydrodynamic torque as
\begin{equation}
    \beta(y) = \frac{m B}{\eta a^3 \dot{\gamma}(y)} = \frac{4 \pi M B}{3 \eta \dot{\gamma} (y)} \label{eq:beta}
\end{equation}
where $\dot{\gamma}(y)$ is the local shear rate of the flow around the particle.
For example, when we assume that the particle magnetization $\mu_0 M = 10^{-3}$ $\rm{T}$ where $\mu_0 = 4\pi \times 10^{-7}$ $\rm{N/A^2}$ is the permeability of free space, particle size $a = 10^{-5}$ $\rm{m}$, water viscosity $\eta = 10^{-3}$ $\rm{Pa \cdot s}$, water density $\rho = 10^3$ $\rm{kg/m^3}$, shear rate $\dot{\gamma} = 10^2$ $\rm{s^{-1}}$ and magnetic field $B = 1.0^{-4} - 10^{-2}$ $\rm{T}$, the particle Reynolds number is $Re \approx 10^{-2}$ and $\beta \approx 10^{0} - 10^2$.

We use the boundary element method \cite{PozrikidisBook, Ishikawa2006, Mitchell2015} to solve for particle trajectory.
When inertial effects are negligible, the flow field $\bm{v}$ of a given point $\bm{x}$ under Stokes flow can be described using a boundary integral formulation \cite{PozrikidisBook}:
$v_i({\bf x}) = v_i^\infty ({\bf x}) - (1/8 \pi \eta) \int_A G_{ij} ({\bf x}, {\bf y}) q_j ({\bf y})dA$
where ${\bf G}$ is the Green's function, $\bf{v}^\infty$ is the background flow, and $\bf{q}$ is the viscous traction acting at a point ${\bf y}$ on the particle surface.
The Blake tensor \cite{Blake1971} is used for the Green's function $G_{ij}$ to account for the walls.
Integrating the traction force ${\bf q}$ on the surface of ellipsoid $A$ gives the hydrodynamic force ${\bf F}_h$ and torque ${\bf T}_h$ acting on the particle. As the system is force- and torque-free, these satisfy
${\bf F}_{h} = \int_A {\bf q} dA = {\bf 0}$,
${\bf T}_{h} + {\bf T}_{m} = \int_A \{ {\bf q} \times ({\bf x} - {\bf x}_0) \} dA  + {\bf m} \times {\bf B} = \bf{0}$,
where ${\bf x}_0$ is the hydrodynamic centre of the particle \cite{KimBook}.
A given surface material point $\bf{x}_s$ on the ellipsoid moves with a velocity ${\bf v} ({\bf x}_s) = {\bf U} + {\bf \Omega} \times ({\bf x}_s - {\bf x}_0)$, where ${\bf U}$, ${\bf \Omega}$ are the translational and rotational velocity of the particle, respectively.
The surface of the ellipsoid is divided into $N_E = 512$ triangular elements and $N_N = 258$ nodes.
The velocities are obtained by solving the dense matrix ${\bf Ax}={\bf b}$ with a known vector ${\bf b}=({\bf v}^\infty, {\bf F}_h, {\bf T}_h)$ and an unknown vector ${\bf x}=({\bf q}, {\bf U}, {\bf \Omega})$, and ${\bf A}$ is a matrix with size $(3N_N + 6)$ based on equations above \cite{Ishikawa2006}.
The particle position is updated using the first-order Euler method with a time step $\dot{\gamma} \Delta t = 0.01$.
The software is written in CUDA and all processes are parallelized \cite{Matsunaga2014}.

First, we show that transverse motion can be manipulated by pinning the rotational motion of the particle in shear flow, $v^\infty_x = \dot{\gamma} y$ ($\dot{\gamma}(y) = \rm{const}$).
The rotational motion of an ellipsoidal particle subjected to shear and a magnetic field was discussed in Ref.~\cite{Almog1995}, in the absence of a wall.
The authors showed that the particle moves in the shear plane for sufficiently large $\beta$ and reaches a stable angle $\phi_p^*$  where it is pinned by the magnetic field.
The general expression for the in-plane rotational velocity is 
\begin{equation}
    \frac{1}{\dot{\gamma}(y)} \dot{\phi}_p = \frac{\beta (y)}{8 \pi} F(\alpha) \sin(\phi_B - \phi_p) - \frac{1}{2} (1 - J(\alpha) \cos 2 \phi_p), \label{eq:d_phi}
\end{equation}
and $\phi_p^*$ is obtained by solving $\dot{\phi}_p = 0$.
Note that the first term of Eq. (\ref{eq:d_phi}) is due to the magnetic torque aligning the particle towards the field orientation $\phi_B$, and the second term is simply Jeffrey's rotation of an ellipsoid in flow \cite{Jeffery1922} with $J(\alpha) = (\alpha^2 - 1)/(\alpha^2 + 1)$ and $F(\alpha) = \frac{3}{2 (\alpha^2 - \alpha^{-2})} \{ \frac{2 \alpha - \alpha^{-1}}{\sqrt{\alpha^2 - 1}} \ln ( \alpha + \sqrt{\alpha^2 - 1})  - 1 \}$ \cite{Koenig1975}.

Figure \ref{fig:figure1}(c) is a schematic of the motion of an ellipsoid in shear flow at different $\beta$, now in the presence of a surface.
In the absence of a magnetic field ($\beta = 0$) the particle rotates and oscillates along the $y$-direction, but has no net displacement along $y$ \cite{Pozrikidis2005a}.
However, when the magnetic field is strong enough to pin the orientation, the ellipsoid either continuously travels upwards or downwards.
The transverse motion can be explained by hydrodynamic interactions between the pinned ellipsoid and the wall. The wall can be considered to act as an image stresslet \cite{Smart1991, Zhao2011, Nix2014}, and the leading order contribution to the lift velocity $U_y$ arises from the stresslet component $S_{yy}$ \cite{Blake1971} evaluated for the stable angle $\phi_p^*$:
 \begin{equation}
    U_y (y, \phi_p^*) = - \frac{9}{64 \pi \eta} P(y) S_{yy} (\phi_p^*) \label{eq:lift}
\end{equation}
where $P(y) = 1/y^2$, and to leading order it is sufficient to approximate $S_{yy}(\phi_p^*)$ by $S_{yy}^\infty (\phi_p^*)$ , which is its value in free space $y \rightarrow \infty$ \cite{KimBook}, given by
\begin{equation}
    \frac{S_{yy}^\infty (\phi_p^*)}{\eta a^3 \dot{\gamma}} = A(\alpha) \sin 2\phi_p^* + B(\alpha) \sin 4\phi_p^* \label{eq:Syy}
\end{equation}
where
$A(\alpha) = \pi\alpha^2 (5 X^M - 5 Z^M + 12 Y^H)/6$,
$B(\alpha) = - 5 \pi \alpha^2 (3 X^M - 4 Y^M + 12 Z^M)/12$
and $X^M, Y^M, Z^M, Y^H$ are shape functions \cite{KimBook, Supplemental} that are only a function of the eccentricity $e = \sqrt{1 - \alpha^{-2}}$.
Since $|A(\alpha)| \gg |B(\alpha)|$ [see inset of Fig.~\ref{fig:figure1}(d)], the stresslet changes its sign only for $\sin 2\phi^\ast_p = 0$ as shown in Fig.~\ref{fig:figure1}(d).
Therefore, the particle moves away from the wall $U_y > 0$ for $\sin 2 \phi_p^* > 0$, while it moves towards the wall $U_y < 0$ for $\sin 2 \phi_p^* < 0$.
Figure \ref{fig:figure1}(e) shows simulation and theoretical results for the lift velocity under strong orientational pinning $\beta = 100$ for different distances of the particle from the surface. Very good agreement is obtained for $y/a \gtrsim 4$.
Deviations occur close to the wall where higher order terms in Eq. (\ref{eq:lift}) play a role.
We also ignored the fact that the stresslet $S_{yy}$ itself is modified due to the presence of the wall.

\begin{figure*}
    \includegraphics[width=2.00\columnwidth]{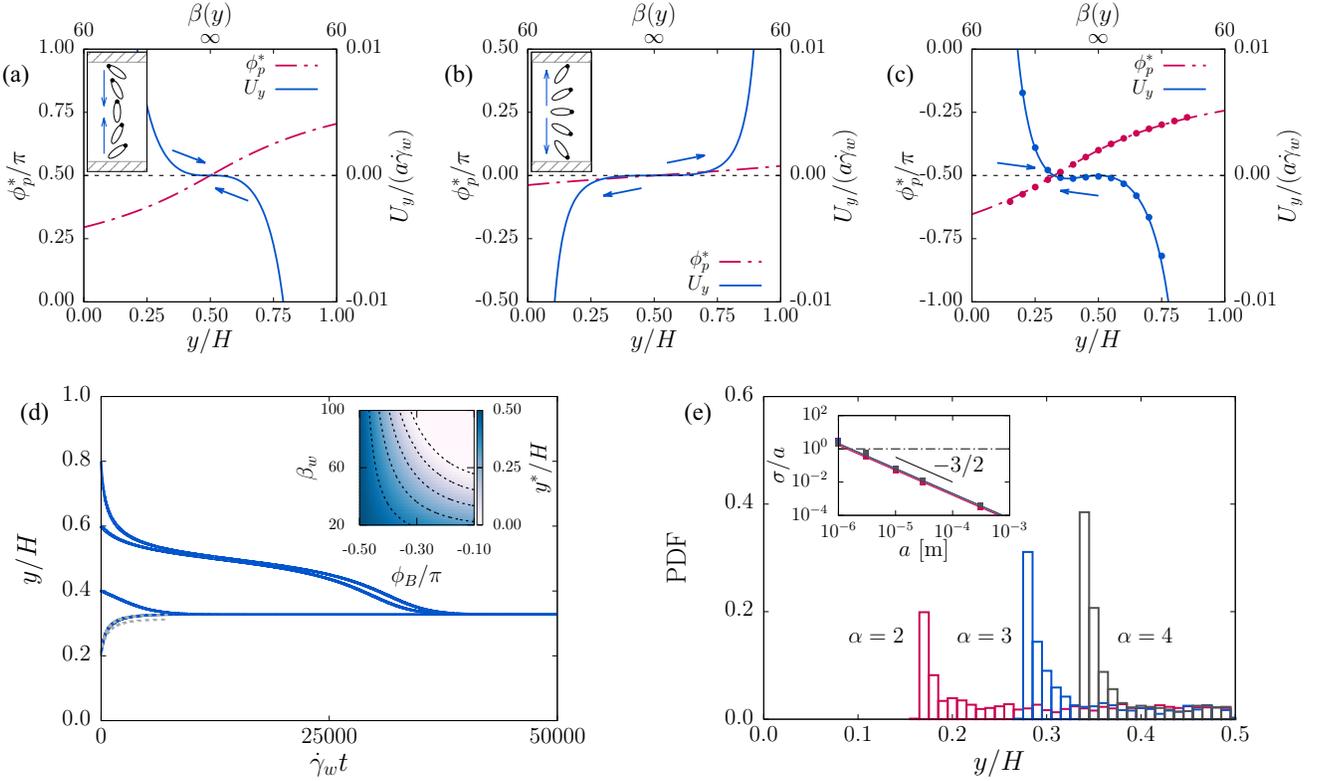}
    \caption{Focusing under Poiseuille flow: (a)-(c) Stable angle $\phi_p^*$ and transverse velocity $U_y$ as a function of the particle ($\alpha = 3$) position $y$ for a magnetic field $\beta_w = 60$ and $\phi_B =$ (a) $\pi/2$, (b) $0$ and (c) $-0.4\pi$ and channel width $H/a = 20$. Lines: prediction from far-field theory; dots in (c): results from boundary element simulations. (d) Time history of particle position $y(t)$ from boundary element simulations under the conditions of (c). The gray dotted curves show trajectories in a rectangular channel of aspect ratio $H_z/H_y=4$ for different initial conditions $z_0=\{0.5H_z, 0.8 H_z\}$. The inset shows stable fixed point $y^*/H$ for ellipsoids ($\alpha=3$) as a function of the magnetic field $\beta_w$ and $\phi_B$ obtained by far-field theory. White lines are isolines for every 0.1. (e) Distribution of focussed particles under Poiseuille flow with $\dot{\gamma}_w = 100$ $\rm{s^{-1}}$ and channel width $H/a = 20$ after 10s, from Brownian dynamics simulations with particle size $a = 10$ $\rm{\mu m}$. Inset: theoretical predictions (lines) and Brownian dynamics simulation (dots) for the standard deviation of the distribution $\sigma/a$. Dotted horizontal line indicates separations $a \gtrsim 2$ $\rm{\mu m}$. \label{fig:figure2}}
\end{figure*}

Next we show that the magnetic particle can be focused to an arbitrary transverse position under Poiseuille flow between two walls.
This geometry is an approximation for high aspect ratio rectangular channels away from side walls. 
The background velocity profile is $v_x^\infty (y) = \dot{\gamma}_w y (H - y)/H$, with $\dot{\gamma}_w$ the shear rate at the wall and $H$ the distance between the two walls [Fig.~\ref{fig:figure1}(b)].
$\beta$ [Eq.~(\ref{eq:beta})] can be locally defined as $\beta (y) = \beta_w/(1 - 2y/H)$ with $\beta_w = mB/(\eta a^3 \dot{\gamma}_w)$ describing the value at the wall. 
Since the rotational motion is much faster than the translational motion, the particle angle $\phi_p$ can be assumed to be quasi-static for a given position $y$.
Hence the far-field approximation of the position-dependent stable angle $\phi_p^* (y)$ follows by simply solving $\dot{\phi}_p = 0$ [Eq. (\ref{eq:d_phi})] at each $y$.
As shown in Fig.~\ref{fig:figure2}(a)-(c), the angle $\phi_p^*$ increases with $y$ because the local vorticity of the Poiseuille flow $\omega_z = \dot{\gamma}_w (2y/H - 1)$ monotonically increases with $y$ while the magnetic contribution is constant throughout the space.
Note that at the channel center $\phi_p^* (H/2) = \phi_B$ since the local shear rate is zero, and hence the angle is only determined by the magnetic torque.
Again, the far-field approximation to the transverse velocity $U_y$ is obtained by stresslet images (\ref{eq:lift}), but with revised position factor $P(y) = (1 - 2y/H)(1/y^2 - 1/(H - y)^2)$ to take into account the effect of two walls.
The velocity $U_y$ is shown in Fig.~\ref{fig:figure2}(a)-(c).
Since $P(y)$ is always positive, the sign of the stresslet $S_{yy}$ alone determines the $y$-directional movement of the particle: equivalently, the stable angle $\phi_p^*$ determines the direction [see Fig.~\ref{fig:figure1}(d)].

A stable fixed point $y^*$, which is determined by $U_y (y^*)= 0$ and $\partial U_y/ \partial y |_{y = y^*} < 0$, is required to focus the particle to a specific position. 
When the magnetic field is applied perpendicular to the flow direction ($\phi_B = \pi/2$) as shown in Fig.~\ref{fig:figure2}(a), the particles are focused to the channel center because the change in sign of $\sin 2\phi_p^* > 0$ $(y < H/2)$ to $\sin 2\phi_p^* < 0$ $(y > H/2)$ gives the conditions for a stable fixed point at $y = H/2$ [inset of Fig.~\ref{fig:figure2}(a)].
This relation is reversed when the magnetic field points in the flow direction ($\phi_B = 0$) [Fig.~\ref{fig:figure2}(b)], and all particles move towards the walls.
In general, a particle position $y^*$ satisfying $\phi_p^* (y^*) = \pm \pi/2$ is a stable fixed point, while a position with $\phi_p^*(y^*) = 0$ or $\pi$ is an unstable fixed point.

Hence by changing the direction of the magnetic field $\phi_B$, it is possible to focus particles to arbitrary target positions.
The stable fixed point $y^*$ can be predicted solving $\dot{\phi}_p = 0$ [Eq. (\ref{eq:d_phi})] for $\phi_p^* = \pm \pi/2$ as
\begin{equation}
    \frac{y^*}{H} (\alpha, \beta_w, \phi_B) = \frac{1}{2} \pm \frac{\beta_w F(\alpha)}{8 \pi (1 + J(\alpha))} \cos \phi_B \label{eq:stable}
\end{equation}
where $J(\alpha)$ and $F(\alpha)$ are defined after Eq.(\ref{eq:d_phi}).
For example, when a field is applied in direction $\phi_B = -0.4\pi$ with $\beta_w = 60$ [Fig.~\ref{fig:figure2}(c)], the stable fixed point of an ellipsoid ($\alpha = 3$) is $y^*/H \approx 0.32$ and the particles are focused to this position. 
This we also confirm by performing boundary element simulations with different initial positions, shown in Fig.~\ref{fig:figure2}(d), and we observe that $U_y(y)$ and  $\phi_p^* (y)$ obtained from simulations qualitatively agree with the far-field results.
Note that the hydrodynamic contributions from the two walls are calculated considering the images of both walls \cite{Blake1971} in the simulation, which is enough for relatively large channels $H/a = 20$, while higher order reflections are required for a much narrower channel \cite{Supplemental, Mathijssen2016}.
Inset of Fig.~\ref{fig:figure2}(d) describes the stable fixed point $y^*/H$ [Eq. (\ref{eq:stable})] for ellipsoids ($\alpha = 3$) with $-\pi/2 < \phi_B < 0$ .
The figure shows $y^*$ shifts toward the bottom wall with increasing $\beta_w$ or $\phi_B$, and the particles can be focused to arbitrary positions in the lower half of the channel ($y/H < 0.5$). 
By symmetry, the particles would be focused to the upper half of the channel for $0 < \phi_B < \pi/2$.
Although the far-field theory predicts that the particles cannot cross the center line because $U_y(H/2) = 0$, in reality they can do so because of their finite size, as confirmed by boundary element simulations [Fig.~\ref{fig:figure2}(d)]. 

To see the effect of side walls, at $z=0$ and $z=H_z$, on the motion of the particles, we extended our simulation scheme to a rectangular channel geometry by using a triangular mesh both for the particles and the walls \cite{Supplemental, Pozrikidis2005b, Hu2012, Mortensen2005}.
When the particles are not too close to the walls  we observe very similar trajectories, focusing points $y^\ast$ and focusing times as  without side walls \cite{Supplemental}  [Fig.~\ref{fig:figure2}(d)]. Moreover  migration in $z$ direction is negligible.

Finally we show how a static magnetic field can be used to separate particles of different aspect ratio $\alpha$ even in the presence of thermal fluctuations.
The particles  are initially uniformly distributed in the lower half of the channel \cite{Squires2005}, and we consider the same magnetic field and channel height as discussed above ($\beta_w=60$, $\phi_B=-0.4\pi$, $H/a=20$), wall-shear rate $\dot{\gamma}_w = 100$ $\rm{s^{-1}}$, and viscosity of water ($\eta=10^{-3}$ $\rm{Pa \cdot s}$).
We use Brownian Dynamics simulations at room temperature, solving the equations
\begin{eqnarray}
   \dot{\mathbf{r}} &=& \mathbf{U}_p +  \mathcal{H} \cdot\boldsymbol{\xi}, \\
   \dot{\mathbf{n}} &=& (\boldsymbol{\Omega}_p + \sqrt{2D_r}\boldsymbol{\xi}^r) \times \mathbf{n}
\end{eqnarray}
for different particle size $a$ and aspect ratio $\alpha$.
Here $\mathbf{U}_p=v_x^\infty \hat{\mathbf{x}} + U_y\hat{\mathbf{y}}$, and $\boldsymbol{\Omega}_p=\Omega_\phi \hat{\boldsymbol{\phi}} + \Omega_\theta \hat{\boldsymbol{\theta}}$
 is the full 3D particle reorientation rate for the particle orientation $\mathbf{n} = (\sin\theta_p\cos\phi_p, \sin\theta_p\sin\phi_p, \cos\theta_p)$
with 
$\Omega_\phi = \dot{\gamma}_w \{\beta(y) F(\alpha) \sin(\phi_B-\phi_p) /(8\pi \sin\theta_p) - (1-J\cos 2 \phi_p)/2 \}$, 
$\Omega_\theta =\dot{\gamma}_w \{\beta(y) F(\alpha) \cos\theta_p \cos(\phi_B-\phi_p)/(8\pi) + J\sin 2 \theta_p\sin 2 \phi_p /4 \}$ \cite{Almog1995}.
$\mathcal{H}$ is calculated from the translational  diffusion tensor $\textbf{\textsf{D}}(\phi_p,\theta_p)=\bar{D}{\textbf{\textsf{1}}} + \frac 1 2 \Delta D \textbf{\textsf{M}}(\phi_p,\theta_p) =\frac 1 2 \mathcal{H} \cdot \mathcal{H}^T$
where $\textbf{\textsf{M}}(\phi_p,\theta_p)$ is a symmetric 3x3 matrix \cite{Supplemental} and $\bar{D} = (D_1 + D_2)/2$, $\Delta D = D_1 - D_2$ where $D_1=k_BT a^{-1} \eta^{-1} K_1(\alpha)$ and $D_2 = k_BT a^{-1} \eta^{-1} K_2(\alpha)$
are the respective longitudinal and transverse diffusion coefficients of an ellipsoid of aspect ratio $\alpha$ with shape functions $K_1(\alpha) > K_2(\alpha)$ \cite{KimBook, Han2006, Supplemental, Cobb2005, Corato2015}.
The rotational diffusion constant $D_r=k_BT a^{-3} \eta^{-1}K_r(\alpha)$ with the shape function $K_r(\alpha)$ \cite{Koenig1975,KimBook,Supplemental}.
The random numbers $\xi_i$ and $\xi_i^r$ model Gaussian white noise with zero mean and $\langle \xi_i \xi_j \rangle=\langle \xi_i^r \xi_j^r \rangle=\delta_{ij}$ ($i=x,y,z$).

Distributions for 1000 particles of size $a = 10$ $\rm{\mu m}$ for $\alpha=\{2,3,4 \}$ after $t=10$s are shown in Fig.~\ref{fig:figure2}(e).
Our results clearly show that particles of different shape can be separated to different target positions $y_{(\alpha)}^*$, given by Eq. (\ref{eq:stable}), by applying a static magnetic field.
50\% of the particles reach the target region $y_{(\alpha)}^* \pm a$ in experimentally feasible times [7s ($\alpha = 4$) to 20s ($\alpha = 2$)] and traveling distances ($<30$ mm).
Note, the focusing times are even smaller for higher confinement \cite{Supplemental}.
Efficient separation is only possible for particles of size $a \gtrsim 2$ $\rm{\mu m}$, where the width of the steady state distribution $\sigma$ \cite{HonerkampBook} is smaller than the distance between two peaks [see inset of Fig.~\ref{fig:figure2}(e)].
We find an approximate analytic expression for $\sigma$ by linearizing the drift velocity around the fixed point $y^*$, $U_y = -k(y - y^*)$ where $k$ only depends on the system parameters \cite{Supplemental}.
We solve for the steady state distribution $p(y)\sim \exp[-V/k_BT]$ where we introduced a potential $V=\gamma_1 k(y-y^*)^2/2 $ with $\gamma_1 = k_BT/ D_1$ which keeps the particle near its target position $y^*$.
Since $k\sim s^{-1}$ and $\gamma_1 \sim \eta a$ we obtain $\sigma/a \sim a^{-3/2}\eta^{-1/2}(k_BT)^{1/2}$.

We have shown that the transverse position of magnetic ellipsoidal particles in microchannel Poiseuille flow can be controlled by a static magnetic field.
This is due to the hydrodynamic interactions of the ellipsoids with the channel walls.
Our method can be used to focus and segregate magnetic particles which is of importance for particle manipulation in lab-on-a-chip devices.

\begin{acknowledgments}
This project has received funding from the European Union's Horizon 2020 research and innovation programme under grant agreement No. 665440 and under the Marie Sklodowska-Curie grant agreement No 653284.
\end{acknowledgments}

\bibliography{reference}

\providecommand{\noopsort}[1]{}\providecommand{\singleletter}[1]{#1}%
\begin{thebibliography}{48}%
\makeatletter
\providecommand \@ifxundefined [1]{%
 \@ifx{#1\undefined}
}%
\providecommand \@ifnum [1]{%
 \ifnum #1\expandafter \@firstoftwo
 \else \expandafter \@secondoftwo
 \fi
}%
\providecommand \@ifx [1]{%
 \ifx #1\expandafter \@firstoftwo
 \else \expandafter \@secondoftwo
 \fi
}%
\providecommand \natexlab [1]{#1}%
\providecommand \enquote  [1]{``#1''}%
\providecommand \bibnamefont  [1]{#1}%
\providecommand \bibfnamefont [1]{#1}%
\providecommand \citenamefont [1]{#1}%
\providecommand \href@noop [0]{\@secondoftwo}%
\providecommand \href [0]{\begingroup \@sanitize@url \@href}%
\providecommand \@href[1]{\@@startlink{#1}\@@href}%
\providecommand \@@href[1]{\endgroup#1\@@endlink}%
\providecommand \@sanitize@url [0]{\catcode `\\12\catcode `\$12\catcode
  `\&12\catcode `\#12\catcode `\^12\catcode `\_12\catcode `\%12\relax}%
\providecommand \@@startlink[1]{}%
\providecommand \@@endlink[0]{}%
\providecommand \url  [0]{\begingroup\@sanitize@url \@url }%
\providecommand \@url [1]{\endgroup\@href {#1}{\urlprefix }}%
\providecommand \urlprefix  [0]{URL }%
\providecommand \Eprint [0]{\href }%
\providecommand \doibase [0]{http://dx.doi.org/}%
\providecommand \selectlanguage [0]{\@gobble}%
\providecommand \bibinfo  [0]{\@secondoftwo}%
\providecommand \bibfield  [0]{\@secondoftwo}%
\providecommand \translation [1]{[#1]}%
\providecommand \BibitemOpen [0]{}%
\providecommand \bibitemStop [0]{}%
\providecommand \bibitemNoStop [0]{.\EOS\space}%
\providecommand \EOS [0]{\spacefactor3000\relax}%
\providecommand \BibitemShut  [1]{\csname bibitem#1\endcsname}%
\let\auto@bib@innerbib\@empty
\bibitem [{\citenamefont {Dreyfus}\ \emph {et~al.}(2003)\citenamefont
  {Dreyfus}, \citenamefont {Tabeling},\ and\ \citenamefont
  {Willaime}}]{Dreyfus2003}%
  \BibitemOpen
  \bibfield  {author} {\bibinfo {author} {\bibfnamefont {R.}~\bibnamefont
  {Dreyfus}}, \bibinfo {author} {\bibfnamefont {P.}~\bibnamefont {Tabeling}}, \
  and\ \bibinfo {author} {\bibfnamefont {H.}~\bibnamefont {Willaime}},\ }\href
  {\doibase 10.1103/PhysRevLett.90.144505} {\bibfield  {journal} {\bibinfo
  {journal} {Phys. Rev. Lett.}\ }\textbf {\bibinfo {volume} {90}},\ \bibinfo
  {pages} {144505} (\bibinfo {year} {2003})}\BibitemShut {NoStop}%
\bibitem [{\citenamefont {Stone}\ \emph {et~al.}(2004)\citenamefont {Stone},
  \citenamefont {Stroock},\ and\ \citenamefont {Ajdari}}]{Stone2004}%
  \BibitemOpen
  \bibfield  {author} {\bibinfo {author} {\bibfnamefont {H.~A.}\ \bibnamefont
  {Stone}}, \bibinfo {author} {\bibfnamefont {A.~D.}\ \bibnamefont {Stroock}},
  \ and\ \bibinfo {author} {\bibfnamefont {A.}~\bibnamefont {Ajdari}},\ }\href
  {\doibase 10.1146/annurev.fluid.36.050802.122124} {\bibfield  {journal}
  {\bibinfo  {journal} {Annu. Rev. Fluid Mech.}\ }\textbf {\bibinfo {volume}
  {36}},\ \bibinfo {pages} {381} (\bibinfo {year} {2004})}\BibitemShut
  {NoStop}%
\bibitem [{\citenamefont {Link}\ \emph {et~al.}(2004)\citenamefont {Link},
  \citenamefont {Anna}, \citenamefont {Weitz},\ and\ \citenamefont
  {Stone}}]{Link2004}%
  \BibitemOpen
  \bibfield  {author} {\bibinfo {author} {\bibfnamefont {D.~R.}\ \bibnamefont
  {Link}}, \bibinfo {author} {\bibfnamefont {S.~L.}\ \bibnamefont {Anna}},
  \bibinfo {author} {\bibfnamefont {D.~A.}\ \bibnamefont {Weitz}}, \ and\
  \bibinfo {author} {\bibfnamefont {H.~A.}\ \bibnamefont {Stone}},\ }\href
  {\doibase 10.1103/PhysRevLett.92.054503} {\bibfield  {journal} {\bibinfo
  {journal} {Phys. Rev. Lett.}\ }\textbf {\bibinfo {volume} {92}},\ \bibinfo
  {pages} {054503} (\bibinfo {year} {2004})}\BibitemShut {NoStop}%
\bibitem [{\citenamefont {Beatus}\ \emph {et~al.}(2006)\citenamefont {Beatus},
  \citenamefont {Tlusty},\ and\ \citenamefont {Bar-Ziv}}]{Beatus2006}%
  \BibitemOpen
  \bibfield  {author} {\bibinfo {author} {\bibfnamefont {T.}~\bibnamefont
  {Beatus}}, \bibinfo {author} {\bibfnamefont {T.}~\bibnamefont {Tlusty}}, \
  and\ \bibinfo {author} {\bibfnamefont {R.}~\bibnamefont {Bar-Ziv}},\ }\href
  {http://dx.doi.org/10.1038/nphys432} {\bibfield  {journal} {\bibinfo
  {journal} {Nat. Phys.}\ }\textbf {\bibinfo {volume} {2}},\ \bibinfo {pages}
  {743} (\bibinfo {year} {2006})}\BibitemShut {NoStop}%
\bibitem [{\citenamefont {Beatus}\ \emph {et~al.}(2012)\citenamefont {Beatus},
  \citenamefont {Bar-Ziv},\ and\ \citenamefont {Tlusty}}]{Beatus2012}%
  \BibitemOpen
  \bibfield  {author} {\bibinfo {author} {\bibfnamefont {T.}~\bibnamefont
  {Beatus}}, \bibinfo {author} {\bibfnamefont {R.~H.}\ \bibnamefont {Bar-Ziv}},
  \ and\ \bibinfo {author} {\bibfnamefont {T.}~\bibnamefont {Tlusty}},\ }\href
  {\doibase https://doi.org/10.1016/j.physrep.2012.02.003} {\bibfield
  {journal} {\bibinfo  {journal} {Phys. Rep.}\ }\textbf {\bibinfo {volume}
  {516}},\ \bibinfo {pages} {103 } (\bibinfo {year} {2012})}\BibitemShut
  {NoStop}%
\bibitem [{\citenamefont {Pamme}(2007)}]{Pamme2007}%
  \BibitemOpen
  \bibfield  {author} {\bibinfo {author} {\bibfnamefont {N.}~\bibnamefont
  {Pamme}},\ }\href {\doibase 10.1039/B712784G} {\bibfield  {journal} {\bibinfo
   {journal} {Lab Chip}\ }\textbf {\bibinfo {volume} {7}},\ \bibinfo {pages}
  {1644} (\bibinfo {year} {2007})}\BibitemShut {NoStop}%
\bibitem [{\citenamefont {Bhagat}\ \emph {et~al.}(2010)\citenamefont {Bhagat},
  \citenamefont {Bow}, \citenamefont {Hou}, \citenamefont {Tan}, \citenamefont
  {Han},\ and\ \citenamefont {Lim}}]{Bhagat2010}%
  \BibitemOpen
  \bibfield  {author} {\bibinfo {author} {\bibfnamefont {A.~A.~S.}\
  \bibnamefont {Bhagat}}, \bibinfo {author} {\bibfnamefont {H.}~\bibnamefont
  {Bow}}, \bibinfo {author} {\bibfnamefont {H.~W.}\ \bibnamefont {Hou}},
  \bibinfo {author} {\bibfnamefont {S.~J.}\ \bibnamefont {Tan}}, \bibinfo
  {author} {\bibfnamefont {J.}~\bibnamefont {Han}}, \ and\ \bibinfo {author}
  {\bibfnamefont {C.~T.}\ \bibnamefont {Lim}},\ }\href {\doibase
  10.1007/s11517-010-0611-4} {\bibfield  {journal} {\bibinfo  {journal} {Med.
  Biol. Eng. Comput.}\ }\textbf {\bibinfo {volume} {48}},\ \bibinfo {pages}
  {999} (\bibinfo {year} {2010})}\BibitemShut {NoStop}%
\bibitem [{\citenamefont {Sajeesh}\ and\ \citenamefont
  {Sen}(2014)}]{Sajeesh2014}%
  \BibitemOpen
  \bibfield  {author} {\bibinfo {author} {\bibfnamefont {P.}~\bibnamefont
  {Sajeesh}}\ and\ \bibinfo {author} {\bibfnamefont {A.~K.}\ \bibnamefont
  {Sen}},\ }\href {\doibase 10.1007/s10404-013-1291-9} {\bibfield  {journal}
  {\bibinfo  {journal} {Microfluid. Nanofluid.}\ }\textbf {\bibinfo {volume}
  {17}},\ \bibinfo {pages} {1} (\bibinfo {year} {2014})}\BibitemShut {NoStop}%
\bibitem [{\citenamefont {Lindner}(2014)}]{Lindner2014}%
  \BibitemOpen
  \bibfield  {author} {\bibinfo {author} {\bibfnamefont {A.}~\bibnamefont
  {Lindner}},\ }\href {\doibase 10.1063/1.4899260} {\bibfield  {journal}
  {\bibinfo  {journal} {Phys. Fluids}\ }\textbf {\bibinfo {volume} {26}},\
  \bibinfo {pages} {101307} (\bibinfo {year} {2014})}\BibitemShut {NoStop}%
\bibitem [{\citenamefont {Das}\ \emph {et~al.}(2015)\citenamefont {Das},
  \citenamefont {Garg}, \citenamefont {Campbell}, \citenamefont {Howse},
  \citenamefont {Sen}, \citenamefont {Velegol}, \citenamefont {Golestanian},\
  and\ \citenamefont {Ebbens}}]{Das2015}%
  \BibitemOpen
  \bibfield  {author} {\bibinfo {author} {\bibfnamefont {S.}~\bibnamefont
  {Das}}, \bibinfo {author} {\bibfnamefont {A.}~\bibnamefont {Garg}}, \bibinfo
  {author} {\bibfnamefont {A.~I.}\ \bibnamefont {Campbell}}, \bibinfo {author}
  {\bibfnamefont {J.}~\bibnamefont {Howse}}, \bibinfo {author} {\bibfnamefont
  {A.}~\bibnamefont {Sen}}, \bibinfo {author} {\bibfnamefont {D.}~\bibnamefont
  {Velegol}}, \bibinfo {author} {\bibfnamefont {R.}~\bibnamefont
  {Golestanian}}, \ and\ \bibinfo {author} {\bibfnamefont {S.~J.}\ \bibnamefont
  {Ebbens}},\ }\href {http://dx.doi.org/10.1038/ncomms9999} {\bibfield
  {journal} {\bibinfo  {journal} {Nat. Commun.}\ }\textbf {\bibinfo {volume}
  {6}},\ \bibinfo {pages} {8999 EP } (\bibinfo {year} {2015})}\BibitemShut
  {NoStop}%
\bibitem [{\citenamefont {Bechinger}\ \emph {et~al.}(2016)\citenamefont
  {Bechinger}, \citenamefont {Di~Leonardo}, \citenamefont {L\"owen},
  \citenamefont {Reichhardt}, \citenamefont {Volpe},\ and\ \citenamefont
  {Volpe}}]{Bechinger2017}%
  \BibitemOpen
  \bibfield  {author} {\bibinfo {author} {\bibfnamefont {C.}~\bibnamefont
  {Bechinger}}, \bibinfo {author} {\bibfnamefont {R.}~\bibnamefont
  {Di~Leonardo}}, \bibinfo {author} {\bibfnamefont {H.}~\bibnamefont
  {L\"owen}}, \bibinfo {author} {\bibfnamefont {C.}~\bibnamefont {Reichhardt}},
  \bibinfo {author} {\bibfnamefont {G.}~\bibnamefont {Volpe}}, \ and\ \bibinfo
  {author} {\bibfnamefont {G.}~\bibnamefont {Volpe}},\ }\href {\doibase
  10.1103/RevModPhys.88.045006} {\bibfield  {journal} {\bibinfo  {journal}
  {Rev. Mod. Phys.}\ }\textbf {\bibinfo {volume} {88}},\ \bibinfo {pages}
  {045006} (\bibinfo {year} {2016})}\BibitemShut {NoStop}%
\bibitem [{\citenamefont {Nott}\ and\ \citenamefont {Brady}(1994)}]{Brady1994}%
  \BibitemOpen
  \bibfield  {author} {\bibinfo {author} {\bibfnamefont {P.~R.}\ \bibnamefont
  {Nott}}\ and\ \bibinfo {author} {\bibfnamefont {J.~F.}\ \bibnamefont
  {Brady}},\ }\href {\doibase 10.1017/S0022112094002326} {\bibfield  {journal}
  {\bibinfo  {journal} {J. Fluid Mech.}\ }\textbf {\bibinfo {volume} {275}},\
  \bibinfo {pages} {157} (\bibinfo {year} {1994})}\BibitemShut {NoStop}%
\bibitem [{\citenamefont {Squires}\ and\ \citenamefont
  {Quake}(2005)}]{Squires2005}%
  \BibitemOpen
  \bibfield  {author} {\bibinfo {author} {\bibfnamefont {T.~M.}\ \bibnamefont
  {Squires}}\ and\ \bibinfo {author} {\bibfnamefont {S.~R.}\ \bibnamefont
  {Quake}},\ }\href@noop {} {\bibfield  {journal} {\bibinfo  {journal} {Rev.
  Mod. Phys.}\ }\textbf {\bibinfo {volume} {77}},\ \bibinfo {pages} {977}
  (\bibinfo {year} {2005})}\BibitemShut {NoStop}%
\bibitem [{\citenamefont {Segre}\ and\ \citenamefont
  {Silberberg}(1961)}]{Segre1961}%
  \BibitemOpen
  \bibfield  {author} {\bibinfo {author} {\bibfnamefont {G.}~\bibnamefont
  {Segre}}\ and\ \bibinfo {author} {\bibfnamefont {A.}~\bibnamefont
  {Silberberg}},\ }\href {http://dx.doi.org/10.1038/189209a0} {\bibfield
  {journal} {\bibinfo  {journal} {Nature}\ }\textbf {\bibinfo {volume} {189}},\
  \bibinfo {pages} {209} (\bibinfo {year} {1961})}\BibitemShut {NoStop}%
\bibitem [{\citenamefont {Di~Carlo}(2009)}]{DiCarlo2009}%
  \BibitemOpen
  \bibfield  {author} {\bibinfo {author} {\bibfnamefont {D.}~\bibnamefont
  {Di~Carlo}},\ }\href {\doibase 10.1039/B912547G} {\bibfield  {journal}
  {\bibinfo  {journal} {Lab Chip}\ }\textbf {\bibinfo {volume} {9}},\ \bibinfo
  {pages} {3038} (\bibinfo {year} {2009})}\BibitemShut {NoStop}%
\bibitem [{\citenamefont {Prohm}\ and\ \citenamefont
  {Stark}(2014)}]{Prohm2014}%
  \BibitemOpen
  \bibfield  {author} {\bibinfo {author} {\bibfnamefont {C.}~\bibnamefont
  {Prohm}}\ and\ \bibinfo {author} {\bibfnamefont {H.}~\bibnamefont {Stark}},\
  }\href {\doibase 10.1039/C4LC00145A} {\bibfield  {journal} {\bibinfo
  {journal} {Lab Chip}\ }\textbf {\bibinfo {volume} {14}},\ \bibinfo {pages}
  {2115} (\bibinfo {year} {2014})}\BibitemShut {NoStop}%
\bibitem [{\citenamefont {Bretherton}(1962)}]{Bretherton1962}%
  \BibitemOpen
  \bibfield  {author} {\bibinfo {author} {\bibfnamefont {F.~P.}\ \bibnamefont
  {Bretherton}},\ }\href@noop {} {\bibfield  {journal} {\bibinfo  {journal} {J.
  Fluid Mech.}\ }\textbf {\bibinfo {volume} {14}},\ \bibinfo {pages} {284}
  (\bibinfo {year} {1962})}\BibitemShut {NoStop}%
\bibitem [{\citenamefont {Goldman}\ \emph {et~al.}(1967)\citenamefont
  {Goldman}, \citenamefont {Cox},\ and\ \citenamefont {Brenner}}]{Goldman1967}%
  \BibitemOpen
  \bibfield  {author} {\bibinfo {author} {\bibfnamefont {A.}~\bibnamefont
  {Goldman}}, \bibinfo {author} {\bibfnamefont {R.}~\bibnamefont {Cox}}, \ and\
  \bibinfo {author} {\bibfnamefont {H.}~\bibnamefont {Brenner}},\ }\href
  {\doibase http://dx.doi.org/10.1016/0009-2509(67)80048-4} {\bibfield
  {journal} {\bibinfo  {journal} {Chem. Eng. Sci.}\ }\textbf {\bibinfo {volume}
  {22}},\ \bibinfo {pages} {653 } (\bibinfo {year} {1967})}\BibitemShut
  {NoStop}%
\bibitem [{\citenamefont {Pozrikidis}(2005{\natexlab{a}})}]{Pozrikidis2005a}%
  \BibitemOpen
  \bibfield  {author} {\bibinfo {author} {\bibfnamefont {C.}~\bibnamefont
  {Pozrikidis}},\ }\href {\doibase 10.1017/S0022112005006117} {\bibfield
  {journal} {\bibinfo  {journal} {J. Fluid Mech.}\ }\textbf {\bibinfo {volume}
  {541}},\ \bibinfo {pages} {105} (\bibinfo {year}
  {2005}{\natexlab{a}})}\BibitemShut {NoStop}%
\bibitem [{\citenamefont {Guazzelli}\ and\ \citenamefont
  {Morris}(2012)}]{GuazzelliBook}%
  \BibitemOpen
  \bibfield  {author} {\bibinfo {author} {\bibfnamefont {E.}~\bibnamefont
  {Guazzelli}}\ and\ \bibinfo {author} {\bibfnamefont {J.~F.}\ \bibnamefont
  {Morris}},\ }\href@noop {} {\emph {\bibinfo {title} {A Physical Introduction
  to Suspension Dynamics}}}\ (\bibinfo  {publisher} {Cambridge University
  Press},\ \bibinfo {year} {2012})\BibitemShut {NoStop}%
\bibitem [{\citenamefont {Yan}\ \emph {et~al.}(2012)\citenamefont {Yan},
  \citenamefont {Bloom}, \citenamefont {Bae}, \citenamefont {Luijten},\ and\
  \citenamefont {Granick}}]{Yan2012}%
  \BibitemOpen
  \bibfield  {author} {\bibinfo {author} {\bibfnamefont {J.}~\bibnamefont
  {Yan}}, \bibinfo {author} {\bibfnamefont {M.}~\bibnamefont {Bloom}}, \bibinfo
  {author} {\bibfnamefont {S.~C.}\ \bibnamefont {Bae}}, \bibinfo {author}
  {\bibfnamefont {E.}~\bibnamefont {Luijten}}, \ and\ \bibinfo {author}
  {\bibfnamefont {S.}~\bibnamefont {Granick}},\ }\href
  {http://dx.doi.org/10.1038/nature11619} {\bibfield  {journal} {\bibinfo
  {journal} {Nature}\ }\textbf {\bibinfo {volume} {491}},\ \bibinfo {pages}
  {578} (\bibinfo {year} {2012})}\BibitemShut {NoStop}%
\bibitem [{\citenamefont {Hejazian}\ \emph {et~al.}(2015)\citenamefont
  {Hejazian}, \citenamefont {Li},\ and\ \citenamefont {Nguyen}}]{Hejazian2015}%
  \BibitemOpen
  \bibfield  {author} {\bibinfo {author} {\bibfnamefont {M.}~\bibnamefont
  {Hejazian}}, \bibinfo {author} {\bibfnamefont {W.}~\bibnamefont {Li}}, \ and\
  \bibinfo {author} {\bibfnamefont {N.-T.}\ \bibnamefont {Nguyen}},\ }\href
  {\doibase 10.1039/C4LC01422G} {\bibfield  {journal} {\bibinfo  {journal} {Lab
  Chip}\ }\textbf {\bibinfo {volume} {15}},\ \bibinfo {pages} {959} (\bibinfo
  {year} {2015})}\BibitemShut {NoStop}%
\bibitem [{\citenamefont {Huang}\ \emph {et~al.}(2017)\citenamefont {Huang},
  \citenamefont {Yang}, \citenamefont {Zhu}, \citenamefont {Qiao},\ and\
  \citenamefont {Zhao}}]{Huang2017}%
  \BibitemOpen
  \bibfield  {author} {\bibinfo {author} {\bibfnamefont {W.}~\bibnamefont
  {Huang}}, \bibinfo {author} {\bibfnamefont {F.}~\bibnamefont {Yang}},
  \bibinfo {author} {\bibfnamefont {L.}~\bibnamefont {Zhu}}, \bibinfo {author}
  {\bibfnamefont {R.}~\bibnamefont {Qiao}}, \ and\ \bibinfo {author}
  {\bibfnamefont {Y.}~\bibnamefont {Zhao}},\ }\href {\doibase
  10.1039/C7SM00488E} {\bibfield  {journal} {\bibinfo  {journal} {Soft Matter}\
  ,\ } (\bibinfo {year} {2017})}\BibitemShut {NoStop}%
\bibitem [{\citenamefont {Peyer}\ \emph {et~al.}(2013)\citenamefont {Peyer},
  \citenamefont {Zhang},\ and\ \citenamefont {Nelson}}]{Peyer2013}%
  \BibitemOpen
  \bibfield  {author} {\bibinfo {author} {\bibfnamefont {K.~E.}\ \bibnamefont
  {Peyer}}, \bibinfo {author} {\bibfnamefont {L.}~\bibnamefont {Zhang}}, \ and\
  \bibinfo {author} {\bibfnamefont {B.~J.}\ \bibnamefont {Nelson}},\ }\href
  {\doibase 10.1039/C2NR32554C} {\bibfield  {journal} {\bibinfo  {journal}
  {Nanoscale}\ }\textbf {\bibinfo {volume} {5}},\ \bibinfo {pages} {1259}
  (\bibinfo {year} {2013})}\BibitemShut {NoStop}%
\bibitem [{\citenamefont {Hamilton}\ \emph {et~al.}(2017)\citenamefont
  {Hamilton}, \citenamefont {Petrov}, \citenamefont {Winlove}, \citenamefont
  {Gilbert}, \citenamefont {Bryan},\ and\ \citenamefont
  {Ogrin}}]{Hamilton2017}%
  \BibitemOpen
  \bibfield  {author} {\bibinfo {author} {\bibfnamefont {J.~K.}\ \bibnamefont
  {Hamilton}}, \bibinfo {author} {\bibfnamefont {P.~G.}\ \bibnamefont
  {Petrov}}, \bibinfo {author} {\bibfnamefont {C.~P.}\ \bibnamefont {Winlove}},
  \bibinfo {author} {\bibfnamefont {A.~D.}\ \bibnamefont {Gilbert}}, \bibinfo
  {author} {\bibfnamefont {M.~T.}\ \bibnamefont {Bryan}}, \ and\ \bibinfo
  {author} {\bibfnamefont {F.~Y.}\ \bibnamefont {Ogrin}},\ }\href
  {http://dx.doi.org/10.1038/srep44142} {\bibfield  {journal} {\bibinfo
  {journal} {Sci. Rep-UK}\ }\textbf {\bibinfo {volume} {7}},\ \bibinfo {pages}
  {44142 EP } (\bibinfo {year} {2017})}\BibitemShut {NoStop}%
\bibitem [{\citenamefont {Erb}\ \emph {et~al.}(2016)\citenamefont {Erb},
  \citenamefont {Martin}, \citenamefont {Soheilian}, \citenamefont {Pan},\ and\
  \citenamefont {Barber}}]{Erb2016}%
  \BibitemOpen
  \bibfield  {author} {\bibinfo {author} {\bibfnamefont {R.~M.}\ \bibnamefont
  {Erb}}, \bibinfo {author} {\bibfnamefont {J.~J.}\ \bibnamefont {Martin}},
  \bibinfo {author} {\bibfnamefont {R.}~\bibnamefont {Soheilian}}, \bibinfo
  {author} {\bibfnamefont {C.}~\bibnamefont {Pan}}, \ and\ \bibinfo {author}
  {\bibfnamefont {J.~R.}\ \bibnamefont {Barber}},\ }\href {\doibase
  10.1002/adfm.201504699} {\bibfield  {journal} {\bibinfo  {journal} {Adv.
  Funct. Mater.}\ }\textbf {\bibinfo {volume} {26}},\ \bibinfo {pages} {3859}
  (\bibinfo {year} {2016})}\BibitemShut {NoStop}%
\bibitem [{\citenamefont {Zhou}\ \emph {et~al.}(2017)\citenamefont {Zhou},
  \citenamefont {Bai},\ and\ \citenamefont {Wang}}]{Zhou2017}%
  \BibitemOpen
  \bibfield  {author} {\bibinfo {author} {\bibfnamefont {R.}~\bibnamefont
  {Zhou}}, \bibinfo {author} {\bibfnamefont {F.}~\bibnamefont {Bai}}, \ and\
  \bibinfo {author} {\bibfnamefont {C.}~\bibnamefont {Wang}},\ }\href {\doibase
  10.1039/C6LC01382A} {\bibfield  {journal} {\bibinfo  {journal} {Lab Chip}\
  }\textbf {\bibinfo {volume} {17}},\ \bibinfo {pages} {401} (\bibinfo {year}
  {2017})}\BibitemShut {NoStop}%
\bibitem [{\citenamefont {Almog}\ and\ \citenamefont
  {Frankel}(1995)}]{Almog1995}%
  \BibitemOpen
  \bibfield  {author} {\bibinfo {author} {\bibfnamefont {Y.}~\bibnamefont
  {Almog}}\ and\ \bibinfo {author} {\bibfnamefont {I.}~\bibnamefont
  {Frankel}},\ }\href {\doibase 10.1017/S0022112095001327} {\bibfield
  {journal} {\bibinfo  {journal} {J. Fluid Mech.}\ }\textbf {\bibinfo {volume}
  {289}},\ \bibinfo {pages} {243} (\bibinfo {year} {1995})}\BibitemShut
  {NoStop}%
\bibitem [{\citenamefont {Pozrikidis}(1992)}]{PozrikidisBook}%
  \BibitemOpen
  \bibfield  {author} {\bibinfo {author} {\bibfnamefont {C.}~\bibnamefont
  {Pozrikidis}},\ }\href@noop {} {\emph {\bibinfo {title} {Boundary integral
  and singularity methods for linearized viscous flow}}}\ (\bibinfo
  {publisher} {Cambridge University Press},\ \bibinfo {year}
  {1992})\BibitemShut {NoStop}%
\bibitem [{\citenamefont {Ishikawa}\ \emph {et~al.}(2006)\citenamefont
  {Ishikawa}, \citenamefont {Simmonds},\ and\ \citenamefont
  {Pedley}}]{Ishikawa2006}%
  \BibitemOpen
  \bibfield  {author} {\bibinfo {author} {\bibfnamefont {T.}~\bibnamefont
  {Ishikawa}}, \bibinfo {author} {\bibfnamefont {M.~P.}\ \bibnamefont
  {Simmonds}}, \ and\ \bibinfo {author} {\bibfnamefont {T.~J.}\ \bibnamefont
  {Pedley}},\ }\href {\doibase 10.1017/S0022112006002631} {\bibfield  {journal}
  {\bibinfo  {journal} {J. Fluid Mech.}\ }\textbf {\bibinfo {volume} {568}},\
  \bibinfo {pages} {119} (\bibinfo {year} {2006})}\BibitemShut {NoStop}%
\bibitem [{\citenamefont {Mitchell}\ and\ \citenamefont
  {Spagnolie}(2015)}]{Mitchell2015}%
  \BibitemOpen
  \bibfield  {author} {\bibinfo {author} {\bibfnamefont {W.~H.}\ \bibnamefont
  {Mitchell}}\ and\ \bibinfo {author} {\bibfnamefont {S.~E.}\ \bibnamefont
  {Spagnolie}},\ }\href {\doibase 10.1017/jfm.2015.222} {\bibfield  {journal}
  {\bibinfo  {journal} {J. Fluid Mech.}\ }\textbf {\bibinfo {volume} {772}},\
  \bibinfo {pages} {600} (\bibinfo {year} {2015})}\BibitemShut {NoStop}%
\bibitem [{\citenamefont {Blake}(1971)}]{Blake1971}%
  \BibitemOpen
  \bibfield  {author} {\bibinfo {author} {\bibfnamefont {J.~R.}\ \bibnamefont
  {Blake}},\ }\href {\doibase 10.1017/S0305004100049902} {\bibfield  {journal}
  {\bibinfo  {journal} {Math. Proc. Cambridge}\ }\textbf {\bibinfo {volume}
  {70}},\ \bibinfo {pages} {303} (\bibinfo {year} {1971})}\BibitemShut
  {NoStop}%
\bibitem [{\citenamefont {Kim}\ and\ \citenamefont {Karrila}(1991)}]{KimBook}%
  \BibitemOpen
  \bibfield  {author} {\bibinfo {author} {\bibfnamefont {S.}~\bibnamefont
  {Kim}}\ and\ \bibinfo {author} {\bibfnamefont {J.~S.}\ \bibnamefont
  {Karrila}},\ }\href@noop {} {\emph {\bibinfo {title} {Microhydrodynamics -
  Principles and Selected Applications}}}\ (\bibinfo  {publisher} {Dover
  Publications, Inc.},\ \bibinfo {year} {1991})\BibitemShut {NoStop}%
\bibitem [{\citenamefont {Matsunaga}\ \emph {et~al.}(2014)\citenamefont
  {Matsunaga}, \citenamefont {Imai}, \citenamefont {Omori}, \citenamefont
  {Ishikawa},\ and\ \citenamefont {Yamaguchi}}]{Matsunaga2014}%
  \BibitemOpen
  \bibfield  {author} {\bibinfo {author} {\bibfnamefont {D.}~\bibnamefont
  {Matsunaga}}, \bibinfo {author} {\bibfnamefont {Y.}~\bibnamefont {Imai}},
  \bibinfo {author} {\bibfnamefont {T.}~\bibnamefont {Omori}}, \bibinfo
  {author} {\bibfnamefont {T.}~\bibnamefont {Ishikawa}}, \ and\ \bibinfo
  {author} {\bibfnamefont {T.}~\bibnamefont {Yamaguchi}},\ }\href@noop {}
  {\bibfield  {journal} {\bibinfo  {journal} {J. Biomech. Sci. Eng.}\ }\textbf
  {\bibinfo {volume} {14}} (\bibinfo {year} {2014})}\BibitemShut {NoStop}%
\bibitem [{\citenamefont {Jeffery}(1922)}]{Jeffery1922}%
  \BibitemOpen
  \bibfield  {author} {\bibinfo {author} {\bibfnamefont {G.~B.}\ \bibnamefont
  {Jeffery}},\ }\href {\doibase 10.1098/rspa.1922.0078} {\bibfield  {journal}
  {\bibinfo  {journal} {P. Roy. Soc. A-Math Phy.}\ }\textbf {\bibinfo {volume}
  {102}},\ \bibinfo {pages} {161} (\bibinfo {year} {1922})}\BibitemShut
  {NoStop}%
\bibitem [{\citenamefont {Koenig}(1975)}]{Koenig1975}%
  \BibitemOpen
  \bibfield  {author} {\bibinfo {author} {\bibfnamefont {S.~H.}\ \bibnamefont
  {Koenig}},\ }\href {\doibase 10.1002/bip.1975.360141115} {\bibfield
  {journal} {\bibinfo  {journal} {Biopolymers}\ }\textbf {\bibinfo {volume}
  {14}},\ \bibinfo {pages} {2421} (\bibinfo {year} {1975})}\BibitemShut
  {NoStop}%
\bibitem [{\citenamefont {Smart}\ and\ \citenamefont
  {Leighton}(1991)}]{Smart1991}%
  \BibitemOpen
  \bibfield  {author} {\bibinfo {author} {\bibfnamefont {J.~R.}\ \bibnamefont
  {Smart}}\ and\ \bibinfo {author} {\bibfnamefont {J.~D.~T.}\ \bibnamefont
  {Leighton}},\ }\href {\doibase 10.1063/1.857856} {\bibfield  {journal}
  {\bibinfo  {journal} {Phys. Fluids}\ }\textbf {\bibinfo {volume} {3}},\
  \bibinfo {pages} {21} (\bibinfo {year} {1991})}\BibitemShut {NoStop}%
\bibitem [{\citenamefont {Zhao}\ \emph {et~al.}(2011)\citenamefont {Zhao},
  \citenamefont {Spann},\ and\ \citenamefont {Shaqfeh}}]{Zhao2011}%
  \BibitemOpen
  \bibfield  {author} {\bibinfo {author} {\bibfnamefont {H.}~\bibnamefont
  {Zhao}}, \bibinfo {author} {\bibfnamefont {A.~P.}\ \bibnamefont {Spann}}, \
  and\ \bibinfo {author} {\bibfnamefont {E.~S.~G.}\ \bibnamefont {Shaqfeh}},\
  }\href {\doibase 10.1063/1.3669440} {\bibfield  {journal} {\bibinfo
  {journal} {Phys. Fluids}\ }\textbf {\bibinfo {volume} {23}},\ \bibinfo {eid}
  {121901} (\bibinfo {year} {2011})}\BibitemShut {NoStop}%
\bibitem [{\citenamefont {Nix}\ \emph {et~al.}(2014)\citenamefont {Nix},
  \citenamefont {Imai}, \citenamefont {Matsunaga}, \citenamefont {Yamaguchi},\
  and\ \citenamefont {Ishikawa}}]{Nix2014}%
  \BibitemOpen
  \bibfield  {author} {\bibinfo {author} {\bibfnamefont {S.}~\bibnamefont
  {Nix}}, \bibinfo {author} {\bibfnamefont {Y.}~\bibnamefont {Imai}}, \bibinfo
  {author} {\bibfnamefont {D.}~\bibnamefont {Matsunaga}}, \bibinfo {author}
  {\bibfnamefont {T.}~\bibnamefont {Yamaguchi}}, \ and\ \bibinfo {author}
  {\bibfnamefont {T.}~\bibnamefont {Ishikawa}},\ }\href {\doibase
  10.1103/PhysRevE.90.043009} {\bibfield  {journal} {\bibinfo  {journal} {Phys.
  Rev. E}\ }\textbf {\bibinfo {volume} {90}},\ \bibinfo {pages} {043009}
  (\bibinfo {year} {2014})}\BibitemShut {NoStop}%
\bibitem [{Sup()}]{Supplemental}%
  \BibitemOpen
  \href@noop {} {\emph {\bibinfo {title} {See Supplemental
  Material}}}\BibitemShut {NoStop}%
\bibitem [{\citenamefont {Mathijssen}\ \emph {et~al.}(2016)\citenamefont
  {Mathijssen}, \citenamefont {Doostmohammadi}, \citenamefont {Yeomans},\ and\
  \citenamefont {Shendruk}}]{Mathijssen2016}%
  \BibitemOpen
  \bibfield  {author} {\bibinfo {author} {\bibfnamefont {A.~J. T.~M.}\
  \bibnamefont {Mathijssen}}, \bibinfo {author} {\bibfnamefont
  {A.}~\bibnamefont {Doostmohammadi}}, \bibinfo {author} {\bibfnamefont
  {J.~M.}\ \bibnamefont {Yeomans}}, \ and\ \bibinfo {author} {\bibfnamefont
  {T.~N.}\ \bibnamefont {Shendruk}},\ }\href {\doibase 10.1017/jfm.2016.479}
  {\bibfield  {journal} {\bibinfo  {journal} {J. Fluid Mech.}\ }\textbf
  {\bibinfo {volume} {806}},\ \bibinfo {pages} {35} (\bibinfo {year}
  {2016})}\BibitemShut {NoStop}%
\bibitem [{\citenamefont {Pozrikidis}(2005{\natexlab{b}})}]{Pozrikidis2005b}%
  \BibitemOpen
  \bibfield  {author} {\bibinfo {author} {\bibfnamefont {C.}~\bibnamefont
  {Pozrikidis}},\ }\href {\doibase 10.1007/s10665-005-5571-6} {\bibfield
  {journal} {\bibinfo  {journal} {J. Eng. Math.}\ }\textbf {\bibinfo {volume}
  {53}},\ \bibinfo {pages} {1} (\bibinfo {year}
  {2005}{\natexlab{b}})}\BibitemShut {NoStop}%
\bibitem [{\citenamefont {Hu}\ \emph {et~al.}(2012)\citenamefont {Hu},
  \citenamefont {Salsac},\ and\ \citenamefont {Barth{\`e}s-Biesel}}]{Hu2012}%
  \BibitemOpen
  \bibfield  {author} {\bibinfo {author} {\bibfnamefont {X.-Q.}\ \bibnamefont
  {Hu}}, \bibinfo {author} {\bibfnamefont {A.-V.}\ \bibnamefont {Salsac}}, \
  and\ \bibinfo {author} {\bibfnamefont {D.}~\bibnamefont
  {Barth{\`e}s-Biesel}},\ }\href {\doibase 10.1017/jfm.2011.462} {\bibfield
  {journal} {\bibinfo  {journal} {J. Fluid Mech.}\ }\textbf {\bibinfo {volume}
  {705}},\ \bibinfo {pages} {176} (\bibinfo {year} {2012})}\BibitemShut
  {NoStop}%
\bibitem [{\citenamefont {Mortensen}\ \emph {et~al.}(2005)\citenamefont
  {Mortensen}, \citenamefont {Okkels},\ and\ \citenamefont
  {Bruus}}]{Mortensen2005}%
  \BibitemOpen
  \bibfield  {author} {\bibinfo {author} {\bibfnamefont {N.~A.}\ \bibnamefont
  {Mortensen}}, \bibinfo {author} {\bibfnamefont {F.}~\bibnamefont {Okkels}}, \
  and\ \bibinfo {author} {\bibfnamefont {H.}~\bibnamefont {Bruus}},\ }\href
  {\doibase 10.1103/PhysRevE.71.057301} {\bibfield  {journal} {\bibinfo
  {journal} {Phys. Rev. E}\ }\textbf {\bibinfo {volume} {71}},\ \bibinfo
  {pages} {057301} (\bibinfo {year} {2005})}\BibitemShut {NoStop}%
\bibitem [{\citenamefont {Han}\ \emph {et~al.}(2006)\citenamefont {Han},
  \citenamefont {Alsayed}, \citenamefont {Nobili}, \citenamefont {Zhang},
  \citenamefont {Lubensky},\ and\ \citenamefont {Yodh}}]{Han2006}%
  \BibitemOpen
  \bibfield  {author} {\bibinfo {author} {\bibfnamefont {Y.}~\bibnamefont
  {Han}}, \bibinfo {author} {\bibfnamefont {A.~M.}\ \bibnamefont {Alsayed}},
  \bibinfo {author} {\bibfnamefont {M.}~\bibnamefont {Nobili}}, \bibinfo
  {author} {\bibfnamefont {J.}~\bibnamefont {Zhang}}, \bibinfo {author}
  {\bibfnamefont {T.~C.}\ \bibnamefont {Lubensky}}, \ and\ \bibinfo {author}
  {\bibfnamefont {A.~G.}\ \bibnamefont {Yodh}},\ }\href@noop {} {\bibfield
  {journal} {\bibinfo  {journal} {Science}\ }\textbf {\bibinfo {volume}
  {314}},\ \bibinfo {pages} {626} (\bibinfo {year} {2006})}\BibitemShut
  {NoStop}%
\bibitem [{\citenamefont {Cobb}\ and\ \citenamefont {Butler}(2005)}]{Cobb2005}%
  \BibitemOpen
  \bibfield  {author} {\bibinfo {author} {\bibfnamefont {P.~D.}\ \bibnamefont
  {Cobb}}\ and\ \bibinfo {author} {\bibfnamefont {J.~E.}\ \bibnamefont
  {Butler}},\ }\href {\doibase 10.1063/1.1997149} {\bibfield  {journal}
  {\bibinfo  {journal} {J. Chem. Phys.}\ }\textbf {\bibinfo {volume} {123}},\
  \bibinfo {pages} {054908} (\bibinfo {year} {2005})}\BibitemShut {NoStop}%
\bibitem [{\citenamefont {Corato}\ \emph {et~al.}(2015)\citenamefont {Corato},
  \citenamefont {Greco}, \citenamefont {D'Avino},\ and\ \citenamefont
  {Maffettone}}]{Corato2015}%
  \BibitemOpen
  \bibfield  {author} {\bibinfo {author} {\bibfnamefont {M.~D.}\ \bibnamefont
  {Corato}}, \bibinfo {author} {\bibfnamefont {F.}~\bibnamefont {Greco}},
  \bibinfo {author} {\bibfnamefont {G.}~\bibnamefont {D'Avino}}, \ and\
  \bibinfo {author} {\bibfnamefont {P.~L.}\ \bibnamefont {Maffettone}},\ }\href
  {\doibase 10.1063/1.4920981} {\bibfield  {journal} {\bibinfo  {journal} {J.
  Chem. Phys.}\ }\textbf {\bibinfo {volume} {142}},\ \bibinfo {pages} {194901}
  (\bibinfo {year} {2015})}\BibitemShut {NoStop}%
\bibitem [{\citenamefont {Honerkamp}(1994)}]{HonerkampBook}%
  \BibitemOpen
  \bibfield  {author} {\bibinfo {author} {\bibfnamefont {J.}~\bibnamefont
  {Honerkamp}},\ }\href@noop {} {\emph {\bibinfo {title} {Stochastic Dynamical
  Systems}}}\ (\bibinfo  {publisher} {VCH, New York},\ \bibinfo {year}
  {1994})\BibitemShut {NoStop}%
\end{thebibliography}%

\pagebreak
\widetext
\begin{center}
Supplemental Material for ``Focusing and sorting of ellipsoidal magnetic particles in microchannels"
\end{center}

\section{Table of shape functions}
\subsection{Stresslet}
Shape functions $X^M, Y^M, Z^H, Y^H$ \cite{KimBook} appear in Eq. (4) are functions of eccentricity $e = \sqrt{1 - \alpha^{-2}}$ as
\begin{eqnarray}
    X^M (e) &=& \frac{8}{15} e^5 \frac{1}{(3 - e^2) L - 6e}, \\
    Y^M (e) &=& \frac{4}{5} e^5 \frac{2e(1 - 2e^2) - (1 - e^2) L}{(2e(2 e^2 - 3)+3(1 - e^2)L)(-2e + (1 + e^2)L)}, \\
    Z^M (e) &=& \frac{16}{5} e^5 \frac{1 - e^2}{3(1 - e^2)^2 L - 2e (3 - 5e^2)}, \\
    Y^H (e) &=& \frac{4}{3} e^5 \frac{1}{-2e +(1 + e^2)L}, \\
    L(e) &=& \ln \left( \frac{1 + e}{1 - e} \right).
\end{eqnarray}

\subsection{Brownian dynamics}
\begin{itemize}
\item{matrix $\textbf{\textsf{M}}$ in translational diffusion tensor}
\begin{equation}
    M_{ij} =
    \begin{bmatrix}
        -1 + 2 \cos^2 \phi_p \sin^2 \theta_p &
        \sin 2 \phi_p \sin^2 \theta_p &
        \sin 2 \theta_p \cos \phi_p \\
        \sin 2 \phi_p \sin^2 \theta_p &
        -1 + 2 \sin^2 \phi_p \sin^2 \theta_p &
        \sin 2\theta_p \sin \phi_p \\ 
        \sin 2 \theta_p \cos \phi_p & 
        \sin 2\theta_p \sin \phi_p &
        \cos 2 \theta_p
    \end{bmatrix}
\end{equation}

\item{shape coefficients \cite{KimBook} for rotational diffusion}
\begin{eqnarray}
    K_1 (\alpha) &=& \frac{1}{6 \pi \alpha^{2/3} X^A (e)} = \frac{1}{6 \pi \alpha^{2/3}} \frac{3(-2e + (1 + e^2)L)}{8e^3} \\
    K_2 (\alpha) &=& \frac{1}{6 \pi \alpha^{2/3} Y^A (e)} = \frac{1}{6 \pi \alpha^{2/3}} \frac{3(2e + (3e^2 - 1)L)}{16e^3} \\
    K_r (\alpha) &=& \frac{1}{8 \pi \alpha^2 X^C (e)} = \frac{1}{8 \pi \alpha^2} \frac{3(2e - (1 - e^2)L)}{e^3(1 - e^2)}
\end{eqnarray}
\end{itemize}

\newpage
\begin{figure}
    \centerline{\includegraphics[width=0.35\columnwidth]{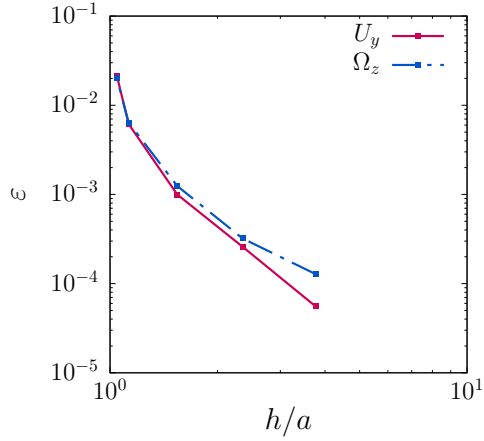}}
    \caption{Relative error $\varepsilon$  of velocity $U_y$ and rotational velocity $\Omega_z$ compared to the analytical solution \cite{Goldman1967}. The variable $h$ is the distance between the sphere center and the plane wall. \label{fig:validation}}
\end{figure}

\section{Validation of boundary element simulation}
In this section, we validate the accuracy of our software by calculating the
 translational velocity $U_y$ and rotational velocity $\Omega_z$ of a sphere close to a wall where analytic expressions exist \cite{Goldman1967}.
The problem setup is the same as the schematic Fig.1(a) in the main text, but a sphere (radius $a$) with no magnetic moment is used for the validation.

Figure \ref{fig:validation} shows that the velocities obtained  have an error of less than $~1\%$ compared to the analytical solution \cite{Goldman1967} for the values,  $h/a \gtrsim 2$, used in the main text. 

\section{Theory for steady state distribution $\sigma$}
In the following we want to calculate how strongly a particle is trapped near a stable point, and compare it with thermal fluctuations.
Therefore we linearize Eq.~(2) of the main text around $\phi_{\pm}^*$ and $y_{\pm}^*$ by taking $\phi=\phi_{\pm}^* + \delta\phi_{\pm}$ and $y=y_{\pm}^*+\delta y_{\pm}$
and find the linearized stable solutions $\Omega_\phi(y_{\pm}^*+\delta y_{\pm},\phi_{\pm}^*+\delta\phi_{\pm})\equiv 0$. Retaining only the linear terms
we obtain the relation 
\begin{equation}
 \delta\phi_{\pm}(y) = \mp\frac{1+J(\alpha)}{2F(\alpha)\beta(y)} - \frac{1}{\tan\phi_B}
\end{equation}
where we have substituted $\delta y_{\pm}$ by $y-y_{\pm}^*$ (note that $y_{\pm}^*$ drops out here).
It can easily be checked that $\delta\phi_{\pm}(y=y_{\pm}^*)=0$.
Now we can calculate the velocity $U_y$ near the fixed point as $U_y(y_{\pm}^* + \delta y_{\pm},\phi_{\pm}^*+\delta\phi_{\pm}) \approx U_y(y,\phi_{\pm}^*+\delta\phi_{\pm}) $,
 and linearize it which results in  $U_y^0 = -k(y-y^*)$ with
\begin{equation}
    k =  8 \dot{\gamma}_w \beta_w \frac{\cos \phi_B}{|\tan\phi_B|}  \frac{a^3 H}{(y_+^*)^2(y_-^*)^2} \frac{ F (\alpha)(A(\alpha) - 2B(\alpha))}{1+J(\alpha)}
\end{equation}
where we skipped the dependence  on $\alpha$ of the parameters $A$,$B$,$F$ and $J$.
Note that $k>0$ and has dimension $s^{-1}$.
We then write the effective Langevin equation for the motion of the ellipsoidal particle in the $y$ direction near the fixed point as
\begin{equation}
    \gamma_1\dot{y} = \gamma_1U_y^0 + \sqrt{2\gamma_1k_BT} = - \frac{dV}{dy} + \sqrt{2\gamma_1k_BT}. 
\label{Eq:L}
\end{equation}
Since the stable orientation is in the $y$ direction 
the particle fluctuations in this direction are around its longitudinal axis, hence we use here its longitudinal friction coefficient $\gamma_1=k_BT/D_1$.
We also introduced a quadratic potential
\begin{equation}
V(y) = \frac{1}{2}\gamma_1k(y-y^*)^2
\end{equation}
defined via $\gamma_1U_y = -dV/dy$.
The stationary solution of Eq.~(\ref{Eq:L}) is a Gaussian distribution \cite{HonerkampBook}
\begin{equation}
p(y) \sim \exp[-\frac{V(y)}{k_BT}] = \exp[-\frac{\gamma_1k(y-y^*)^2}{2 k_BT}] 
\end{equation}
with  standard deviation $\sigma = \sqrt{k_BT/\gamma_1k}$.
Since $\gamma_1\sim \eta\ a$, $k \sim \dot{\gamma}_w$ we get the scaling $\sigma/a \sim a^{-3/2}\eta^{-1/2}(k_BT)^{1/2}$,
see inset of Fig.~2(f) of the main text.

\begin{figure}
    \centerline{\includegraphics[width=0.35\textwidth]{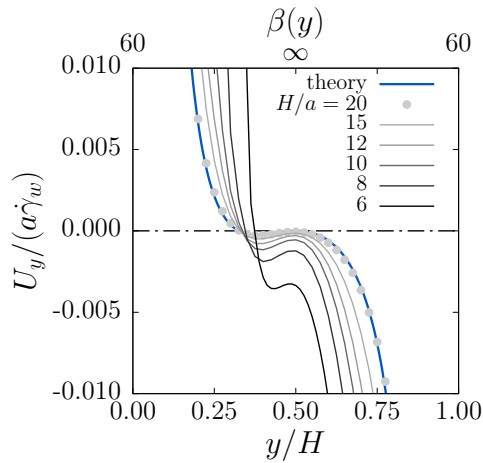}}
    \caption{Effect of confinement $H/a$ on the focusing velocity $U_y$. Conditions are same as Fig. 2(c) in the main text ($\alpha = 3$, $\beta_w = 60$ and $\phi_B = -0.4\pi$). \label{fig:confinement}}
\end{figure}

\section{Effect of confinement}
Figure \ref{fig:confinement} shows the focusing velocity $U_y$ between two meshed walls, which take into account the higher order reflections.
If the distance between the two walls takes the value $H/a = 20$ used in the main text, the far-field theory matches the simulation well. 
Hence this also validates the method of using two meshed walls.
When the distance $H$ is decreased, the velocity starts to deviate from the far-field theory, and the true velocity is faster than the prediction.
Also note that the particle can easily cross the center-line for higher confinement.
The focusing point shifts slightly toward the channel center $y/H = 0.5$, but this effect is not large, in particular for $H/a>10$.

\paragraph{Method}
The numerical method used here is based on previous studies \cite{Hu2012, Pozrikidis2005}, with the confinement effect taken into account by using two meshed walls. Contrary to the mesh on the particle, no-slip boundary conditions are applied for the walls. The dimensions of the two walls ($12.5a \times 12.5a$ - $50.0a \times 50.0a$) and the mesh size ($0.25a-1.00a$ ) were changed depending on the confinement distance $H$.

\begin{figure}
    \begin{minipage}{0.490 \columnwidth}
        \centerline{
            \begin{overpic}[width=\columnwidth]{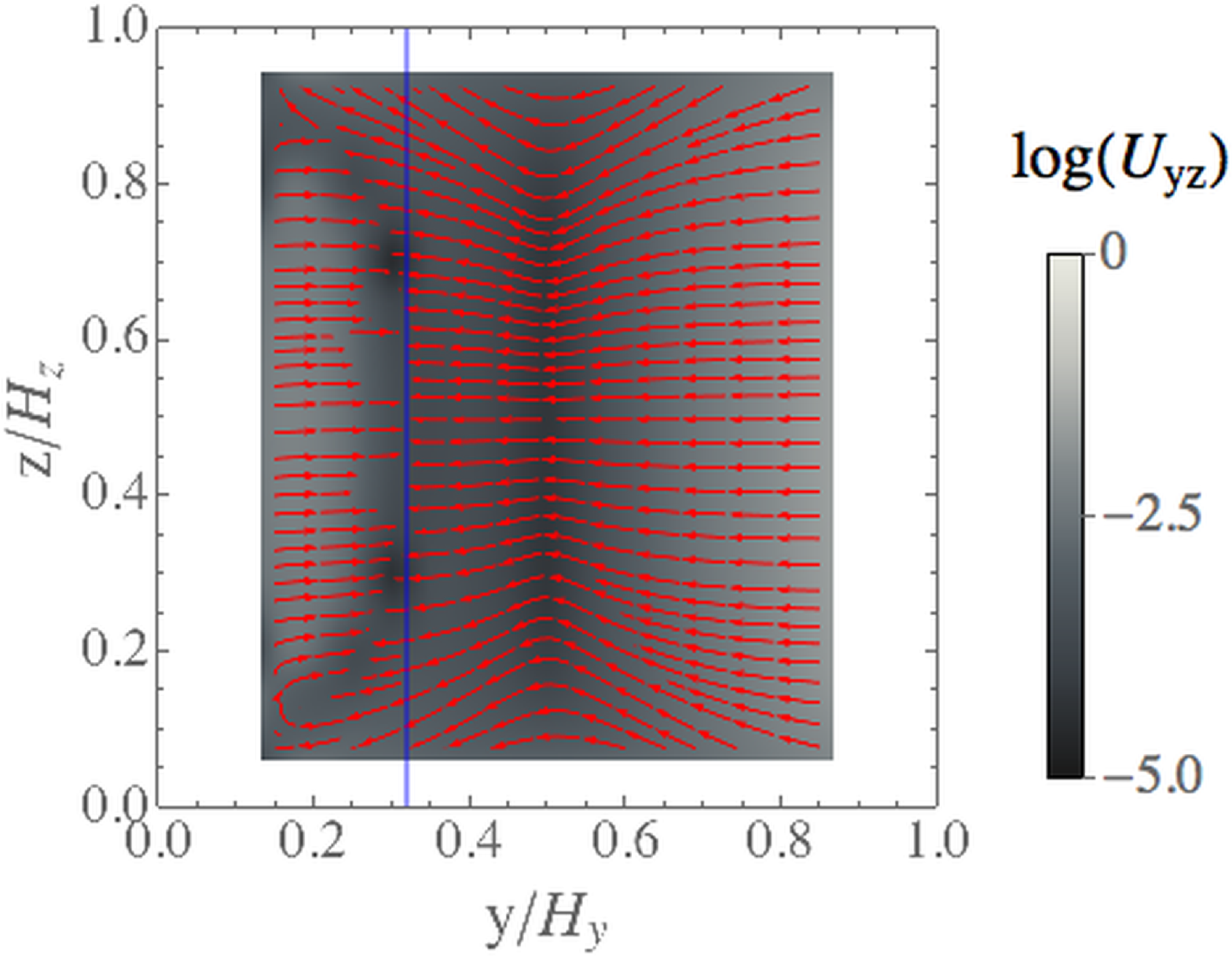}
                \put(0,70){(a)}
            \end{overpic}
        }
    \end{minipage}
    \begin{minipage}{0.490 \columnwidth}
        \centerline{
            \begin{overpic}[width=\columnwidth]{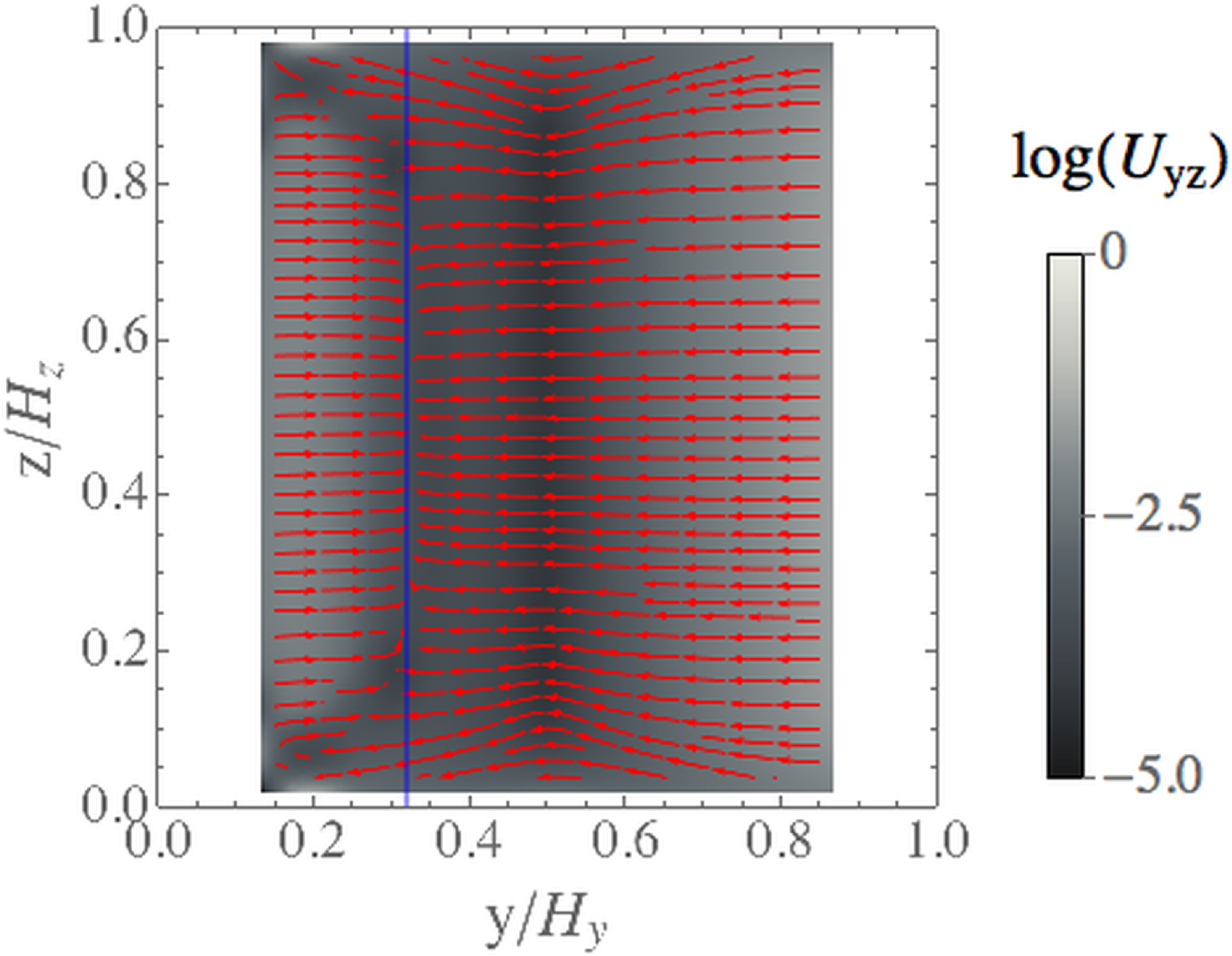}
                \put(0,70){(b)}
            \end{overpic}
        }
    \end{minipage}
    \caption{
        Stream lines (projected to the channel cross-section) of the particles under rectangular channel for two different channel aspect ratios: $H_z/H_y =$ (a) 2 and (b) 4.
        The flow is in $+x$-direction, and the external magnetic field is $\bm{B} = (B \cos \phi_B, B \sin \phi_B, 0)$. Gray scale contours show the amplitude of the in-plane velocity $U_{yz} = \sqrt{U_y^2 + U_z^2}$. Conditions are the same as Fig. 2(c) in the main text ($H_y/a = 20$, $\alpha = 3$, $\beta_w = 60$ and $\phi_B = -0.4 \pi$). Blue lines show the focusing region for two infinite planar walls ($H_z \to \infty$), as discussed in the main text. \label{fig:rectangular}
        }
\end{figure}

\section{Effect of side walls: focusing in a rectangular channel}
Finally, we show particle focusing in a rectangular channel (width $H_y$, height $H_z$).
Figure \ref{fig:rectangular} shows cross-sectional stream lines for two different channel aspect ratios ($H_z/H_y=2$ and $4$).
In the case of two infinite planar walls, the particle is focused to $y^*/H_y = 0.32$ as discussed in the main text.
If the channel is rectangular, most  particles still focus to $y^*/H_y = 0.32$. However particles that are located distances less than $\approx 10a$ away from a side wall reach a curved focusing region at $y^*/H_y = 0.1-0.3$ or the top ($z=H_z$) or bottom ($z=0$) walls.
This effect can be controlled by increasing the channel aspect ratio $H_z/H_y$. 

\paragraph{Method}
We conducted full 3D simulations based on a method used in previous studies \cite{Hu2012, Pozrikidis2005}.
The boundary integral formulation in this system is 
\begin{equation}
    \bm{v}(\bm{x}) = \bm{v}^\infty (\bm{x})
                   - \frac{1}{8 \pi \eta} \left\{ \int_P \bm{G} \cdot \bm{q} dS(\bm{y})
                   + \int_W \bm{G} \cdot \bm{q} dS (\bm{y})
                   + \Delta P \int_O \bm{G} \cdot \bm{n} dS (\bm{y})\right\}
\end{equation}
where the notations $P, W, O$ describe integrals over the ellipsoidal particle (P), side walls (W) and outlet cap (O) respectively.
$\bm{v}^\infty$ is the Poiseuille velocity profile inside a rectangular channel \cite{Mortensen2005}, $\bm{n}$ is the normal vector pointing into the channel, and $\Delta P$ is a additional pressure drop due to the presence of the particle:
\begin{equation}
    \Delta P = \frac{1}{Q} \int_P \bm{v}^\infty(\bm{x}) \cdot \bm{q} dS(\bm{x})
\end{equation}
where $Q$ is the flow rate.
The characteristic shear rate $\dot{\gamma}_w$ is defined at the wall in the two-wall system, while the characteristic shear rate at $(y/H_y, z/H_z) = (0, 0.5)$ is used in the rectangular channel.
The channel length $100a$ and mesh size $2a$ are used in the simulation.
In order to avoid error arising from the coarse mesh, we kept the distance between particle and walls larger than $3a$.
The accuracy of the simulation was validated in a cylindrical channel by comparing with the analytical solution (data not shown). 

\end{document}